\documentclass[12pt, letterpaper]{article}
\usepackage[titletoc,title]{appendix}
\usepackage{color}
\usepackage{booktabs}
\usepackage[usenames,dvipsnames,svgnames,table]{xcolor}
\definecolor{dark-red}{rgb}{0.75,0.10,0.10}
\definecolor{bluish}{rgb}{0.05,0.05,0.85}

\usepackage[margin=1in]{geometry}
\usepackage[linkcolor=blue,
            colorlinks=true,
            urlcolor=blue,
            pdfstartview={XYZ null null 1.00},
            pdfpagemode=UseNone,
            citecolor={bluish},
            pdftitle={Pwned}]{hyperref}

\usepackage{natbib}
\usepackage{float}

\usepackage{geometry} 
\usepackage{graphicx}                
\usepackage{amsfonts,amssymb,amsbsy}
\usepackage{amsxtra}
\usepackage{verbatim}
\setcitestyle{round,semicolon,aysep={},yysep={;}}
\usepackage{setspace}            
\usepackage{sectsty}             
\usepackage{pdflscape}
\usepackage{fancyhdr}            
\usepackage{url}                     
\usepackage{fullpage}       
\usepackage{multirow}
\usepackage{rotating}
\usepackage[T1]{fontenc}
\usepackage[bitstream-charter]{mathdesign}

\usepackage{chngcntr}
\usepackage{booktabs}
\usepackage{longtable}

\usepackage{afterpage}

\makeatother

\usepackage[hang, font=small,skip=0pt]{caption}
\usepackage{subcaption}

\makeatletter
\providecommand\phantomcaption{\caption@refstepcounter\@captype}
\makeatother

\title{Pwned: How Often Are Americans' Online Accounts Breached?\footnote{Data and scripts behind the analysis presented here can be downloaded from \url{https://github.com/themains/pwned}.
}}

\author{Ken Cor\thanks{Ken can be reached at: \href{mailto:mcor@ualberta.ca}{\texttt{mcor@ualberta.ca}}} \and Gaurav Sood\thanks{Gaurav can be reached at: \href{gsood07@gmail.com}{\texttt{gsood07@gmail.com}}}}

\begin{document}
\maketitle
\thispagestyle{empty}

\begin{abstract}
News about massive online breaches is increasingly common. But there has been little good data on how exposed people are because of these breaches. We combine data from a large, representative sample of adult Americans (n = 5,000) with data from \textit{Have I Been Pwned} to estimate the lower bound of the average number of breached online accounts per person. We find that at least 82.84\% of Americans have had their accounts breached. And, on average, Americans' accounts have been breached at least three times. The better educated, the middle-aged, women, and Whites are more likely to have had their accounts breached than the complementary groups.
\end{abstract}
\clearpage

\doublespacing

On the Internet, nobody knows you're a dog. So the adage goes. But increasingly, others know that you like dog food and hate cats. Many of us have made our peace with this new reality. A slew of massive account breaches in recent years \citep{mccandless},\footnote{On September 22, 2016, for instance, Yahoo! revealed that 500M accounts had been compromised in a breach \citep{fiegerman_2016}. Less than three months later, on December 14, 2016, Yahoo! announced that data had been stolen from nearly 1B user accounts in a different breach \citep{newman_2016}. In all, Wikipedia lists 272 separate breaches between 2004 and 2018 (see \href{https://en.wikipedia.org/wiki/List_of_data_breaches}{here}).} however, threaten to pull the rug under all illusions of anonymity. n this note, we shed light on this threat. Using a unique dataset, we estimate the \textit{lower bound} of the average number of breached online accounts per person.

To answer how exposed Americans are due to online breaches, we merge data from a large representative sample from YouGov (n = 5,000) with data from \href{https://haveibeenpwned.com}{Have I Been Pwned} (HIBP). We check whether the email associated with the YouGov account is part of the 293 public breaches cataloged by HIBP.

We find that nearly 83\% of Americans' accounts have been breached at least once. In total, the 5,000 email accounts on file are associated with 14,979 breaches. Or, on average, people are part of three breaches. This number, though, is the lower bound. People generally have more than one email, and we only use one here. And HIBP only catalogs data on a tiny chunk of the total breaches, and only provides data on breaches that do not harm the reputation of the person via its API. 

We also study how exposure to breaches varies by socio-economic factors. Contrary to worries about the digital divide, we find that the kinds of people who are most likely to use online services---the better educated, Whites, etc.---are generally the most exposed. This finding is consistent with \citet{laohaprapanon2018domain}, who find that the better educated, people with higher incomes, and racial majorities spend a \textit{smaller} proportion of time spent online on problematic sites, but because they are online more often, they end up visiting more such sites.

\section*{Data}
In July 2018, YouGov drew a nationally representative sample of 5,000 adult Americans. YouGov drawing a sample in the following way:  it starts by taking a random sample of a high-quality sample of American adults, e.g., Current Population Survey, and then finds people on its panel that match the drawn sample most closely \citep[for more details, see][]{rivers}. Some research suggests that the quality of samples drawn by YouGov is comparable to those drawn using probability sampling \citep{ansolabehere2014does}. However, the sample that YouGov drew here is different from its traditional survey sample in one key aspect. Non-response bias in our sample is 0 because YouGov did not have to send out surveys; it used the emails associated with the accounts to collect the data. (YouGov never shared the emails with us.) Table~\ref{table:yg_dat} presents the marginals on key socio-demographic variables (see \href{https://github.com/themains/pwned/tree/master/data/}{https://github.com/themains/pwned/tree/master/data/} for the codebook). 

\begin{table}[h!]
\centering
\caption{YouGov Sample Marginals.}
\begin{tabular}{ l c }
\hline    
 & proportion \\
\hline
race   & \\
\hspace{2mm}white            &  .67\\
\hspace{2mm}hispanic/latino  &  .13\\
\hspace{2mm}black            &  .12\\
\hspace{2mm}asian            &  .03\\
\hspace{2mm}middle eastern   &  .02\\
\hspace{2mm}mixed race       &  .01\\
\hspace{2mm}native american  &  .01\\
\hspace{2mm}other            &  .00\\
& \\
sex & \\
\hspace{2mm}female           &  .54\\
& \\
age & \\
\hspace{2mm}(18, 25]     & .09 \\
\hspace{2mm}(25, 35]     & .19 \\
\hspace{2mm}(35, 50]     & .26 \\
\hspace{2mm}(50, 65]     & .28 \\
\hspace{2mm}(65, 100]    & .18 \\
& \\
education & \\
\hspace{2mm}no hs                 &   .06\\
\hspace{2mm}hs grad.            &     .32\\
\hspace{2mm}some college        &     .20\\
\hspace{2mm}2-year college degree &   .11\\
\hspace{2mm}4-year college degree &   .19\\
\hspace{2mm}postgrad degree       &   .11\\
\hline
\end{tabular}
\label{table:yg_dat}
\end{table}

After drawing the sample, YouGov used the emails associated with the accounts to query the \url{haveibeenpwned.com} (HIBP) API. HIBP is a non-profit clearinghouse of information about online account breaches. HIBP's stated aim is to provide a way for people to check if they are at risk from online breaches. It currently carries data from 293 breaches covering 278 unique domains and 5,235,843,322 accounts, including data from prominent breaches like the two Yahoo! breaches covering nearly 1.5 billion accounts. The HIBP data are not comprehensive. Security researchers believe that there are many more breaches that the companies are unaware of and at least a few cases where a company doesn't share information about the breach even when it knows. HIBP also refuses to provide data on sensitive breaches---breaches from accounts where a person's inclusion may adversely affect them---from their public API.\footnote{HIBP website notes that it does not share whether or not an account has been part of the breach at ``Adult Friend Finder, Ashley Madison, Beautiful People, Bestialitysextaboo, Brazzers, CrimeAgency vBulletin Hacks, Fling, Florida Virtual School, Freedom Hosting II, Fridae, Fur Affinity, HongFire, Mate1.com, Muslim Match, Naughty America, Non Nude Girls, Rosebutt Board, The Candid Board, The Fappening, xHamster and 1 more.''} So data from HIBP only gives us a lower bound.

HIBP provides an easy way to query all the accounts with a particular email ID have been breached. For instance, to check which accounts associated with the email \url{gsood07@gmail.com} (one of the authors' email) have been breached, you can query \url{https://haveibeenpwned.com/api/v2/breachedaccount/gsood07@gmail.com}. This method gives us all the breaches that HIBP data on for all the email IDs associated with each of the 5,000 profiles. But our YouGov sample provides only one email ID. People often have multiple. So that is another reason why all we get from this data is an absolute lower bound. The actual number is likely much higher than the number we obtain here.

With each request, HIBP returns some metadata on the kind of breaches. (See the \href{https://github.com/themains/pwned/blob/master/data/hibp\_codebook.xlsx}{codebook} for details about all the data that it returns.) Two pieces of information are material here. HIBP catalogs each breach as verified or unverified. And it defines unverified breaches as breaches whose ``legitimacy'' it cannot ``establish beyond reasonable doubt.'' HIBP includes these unverified breaches because ``they still contain personal information about individuals who want to understand their exposure on the web.'' The other material column that HIBP returns relates to whether a breach is part of a ``spam list.'' HIBP defines \texttt{SpamList} as cases where ``large volumes of personal data are found being utilised for the purposes of sending targeted spam.'' HIBP adds, ``This often includes many of the same attributes frequently found in data breaches such as names, addresses, phones numbers and dates of birth. The lists are often aggregated from multiple sources, frequently by eliciting personal information from people with the promise of a monetary reward.'' And the reason HIBP includes these data is: ``whilst the data may not have been sourced from a breached system, the personal nature of the information and the fact that it's redistributed in this fashion unbeknownst to the owners warrants inclusion here.''

\section*{Results}
At least 82.84\% Americans' accounts have been breached at least once. In total, the 5,000 email accounts on file are associated with 14,979 breaches. Or on average, there are three breaches per person. The median is also 3.

The relationship between how frequently emails are part of breaches and socio-economic factors suggests that on the whole who are more likely to use online services are more likely to have had their accounts breached. (See SI \ref{fig:si_tables} and SI \ref{fig:si_figs} for corresponding regressions and figures illustrating group-wise means along with the 95\% confidence intervals.) Women's accounts are 1.2 times more likely to be breached than men's (see Table ~\ref{table:socdem_dat} and \ref{tab:sex_breaches}; $p < .05$). Analyzing breaches by race, African Americans' and Whites' accounts are most frequently breached. The mean number of breaches their email is part of is 3.12 and 3.16 for African Americans and Whites respectively. For Hispanics/Latinos, the corresponding number is 2.5 (see \ref{tab:race_breaches}; $p < .05$). And for Asians, the mean is 2.82. 

\begin{table}[!htb]
\centering
\caption{Frequency of Account Breaches By Socio-economic Factors} 
\label{table:socdem_dat}
\begingroup\small
\begin{tabular}{lrr}
  \hline
 & mean & se \\ 
  \hline
Age &  &  \\ 
  (18,25] & 1.96 & 0.10 \\ 
  (25,35] & 3.12 & 0.09 \\ 
  (35,50] & 3.34 & 0.08 \\ 
  (50,65] & 3.29 & 0.07 \\ 
  (65,100] & 2.95 & 0.07 \\ 
  Missing & 1.19 & 0.16 \\ 
   &  &  \\ 
  Education &  &  \\ 
  No HS & 2.35 & 0.12 \\ 
  HS Grad. & 2.89 & 0.06 \\ 
  Some College & 3.04 & 0.09 \\ 
  2-year College Degree & 3.07 & 0.10 \\ 
  4-year College Degree & 3.22 & 0.09 \\ 
  Postgrad Degree & 3.20 & 0.11 \\ 
   &  &  \\ 
  Sex &  &  \\ 
  Female & 3.17 & 0.05 \\ 
  Male & 2.82 & 0.05 \\ 
   &  &  \\ 
  Race &  &  \\ 
  White & 3.12 & 0.05 \\ 
  Black & 3.16 & 0.11 \\ 
  Hispanic/Latino & 2.50 & 0.08 \\ 
  Asian & 2.82 & 0.21 \\ 
  Native American & 2.96 & 0.26 \\ 
  Middle Eastern & 2.66 & 0.24 \\ 
  Mixed Race & 2.45 & 0.22 \\ 
  Other & 2.92 & 1.32 \\ 
   \hline
\end{tabular}
\endgroup
\end{table}

Looking at education, the relationship is roughly monotonic, with the mean number of breaches increasing with education (see \ref{tab:educ_breaches} and \ref{fig:educ_breaches}). The average number of breaches people with no HS are part of is just 2.35. Compare this to postgraduates, with a mean of 3.20 or 1.3 times as likely.

Lastly, for age, the relationship is curvilinear (see \ref{tab:age_breaches}), with young people's and seniors' accounts least likely to be breached, and middle-aged adults' accounts most likely to be breached, though as the loess illustrates (see \ref{fig:age_breaches}), the relationship is modest.

To assess the source of the exposure, we checked the source of the 14,979 breaches. The 14,979 breaches stemmed from 156 different sites, but there was a sharp skew with 21 sites with more than 100 breaches alone accounting for 11,783 of the breaches. Table ~\ref{table:domain_dat} lists the 21 sites. Prominent websites like \url{linkedin.com}, \url{adobe.com}, \url{dropbox.com}, \url{lastfm.com}, among others feature on the list. 

\begin{table}[h!]
\centering
\caption{Most Frequently Implicated Domains.}
\begin{tabular}{ l c }
\hline    
domain name & n \\
\hline
rivercitymediaonline.com &   2,913 \\
linkedin.com             &   1,089 \\
modbsolutions.com        &   1,067\\
myspace.com              &   1,059\\
data4marketers.com       &    996\\
cashcrate.com            &    856\\
adobe.com                &    609\\
disqus.com               &    570\\
ticketfly.com            &    393\\
tumblr.com               &    340\\
dropbox.com              &    288\\
dailymotion.com          &    255\\
last.fm                  &    248\\
evony.com                &    171\\
clixsense.com            &    150\\
cafemom.com              &    145\\
imesh.com                &    144\\
kickstarter.com          &    140\\
edmodo.com               &    130\\
zomato.com               &    112\\
neopets.com              &    108\\
\hline
\end{tabular}
\label{table:domain_dat}
\end{table}

The analysis until now hasn't distinguished between different kinds of breaches. So next, we shed light on the type of breaches.  Of the 15,837 breaches, 14,979 or 94.58\% were part of verified breaches. And about a third of the 15,837 breaches are categorized as \texttt{SpamList}. In all, we have 10,188 breaches that are verified and not categorized as \texttt{SpamList}. We focus our attention on these breaches, checking whether the relationship with socio-economic variables we see above hold in this smaller subset. When we look at education, the pattern holds up. Once again, the number of breached accounts per person for people with a college degree or more is higher than for people who only got as far as high school (see Table \ref{table:socdem_verified_dat}). Moving to gender, the pattern is more attenuated with women just nudging ahead of men (mean for women and men is 2.15 and 2.05 respectively).  The general pattern for age remains roughly similar to what we saw above, with the middle-aged more likely to have their accounts breached compared to people younger than 25 and older than 65. Breaking down by race, we see some interesting changes. Asians join Whites near the top of the pile, with means of about 2.2. Accounts of Hispanics or Latinos are less likely to be there in verified non-spam-list breaches (mean = 1.73; $p < .05$). The big relative change is for Blacks and that suggests that they are likelier to be part of unverified, spam-list breaches.

\begin{table}[!htb]
\centering
\caption{Frequency of Verified, Non-SpamList Account Breaches By Socioeconomic Factors.} 
\label{table:socdem_verified_dat}
\begingroup\small
\begin{tabular}{lrr}
  \hline
 & mean & se \\ 
  \hline
Age &  &  \\ 
  (18,25] & 1.63 & 0.10 \\ 
  (25,35] & 2.44 & 0.08 \\ 
  (35,50] & 2.37 & 0.07 \\ 
  (50,65] & 2.16 & 0.06 \\ 
  (65,100] & 1.78 & 0.05 \\ 
  Missing & 0.91 & 0.13 \\ 
   &  &  \\ 
  Education &  &  \\ 
  No HS & 1.53 & 0.09 \\ 
  HS Grad. & 1.91 & 0.05 \\ 
  Some College & 2.22 & 0.08 \\ 
  2-year College Degree & 2.10 & 0.08 \\ 
  4-year College Degree & 2.37 & 0.08 \\ 
  Postgrad Degree & 2.30 & 0.08 \\ 
   &  &  \\ 
  Sex &  &  \\ 
  Female & 2.15 & 0.04 \\ 
  Male & 2.05 & 0.05 \\ 
   &  &  \\ 
  Race &  &  \\ 
  White & 2.21 & 0.04 \\ 
  Black & 2.03 & 0.08 \\ 
  Hispanic/Latino & 1.73 & 0.07 \\ 
  Asian & 2.16 & 0.18 \\ 
  Native American & 1.85 & 0.18 \\ 
  Middle Eastern & 2.05 & 0.21 \\ 
  Mixed Race & 1.70 & 0.19 \\ 
  Other & 2.69 & 1.19 \\ 
   \hline
\end{tabular}
\endgroup
\end{table}

\section*{Conclusion}
At least 83\% of Americans' accounts have been breached at least once. And on average, people's accounts have been breached thrice. This is a lower bound for three reasons. First, not all breaches are made public. Second, HIBP doesn't allow access to data on sensitive breaches---breached online accounts on services that may have reputational consequences for people---via its public API. Third, many Americans have multiple email accounts. We only had one email ID per person. 

And generally speaking, the people most at risk are those who are the likeliest to use online services---the better educated, Whites, etc. This runs counter to the traditional narrative about the digital divide, which is that people from lower socioeconomic groups are worse-off. Sometimes, the people holding the shorter end of the stick are those who use online services more.

\clearpage
\bibliographystyle{apsr}
\bibliography{pwned}

\clearpage
\appendix
\renewcommand{\thesection}{SI \arabic{section}}
\setcounter{table}{0}\renewcommand\thetable{\thesection.\arabic{table}}  
\setcounter{figure}{0}\renewcommand\thefigure{\thesection.\arabic{figure}}
\counterwithin{figure}{section}

\section{Supporting Information}

\subsection{Tables}
\label{fig:si_tables}

\begin{table}[!htbp] \centering 
  \caption{Number of Breaches by Race/Ethnicity} 
  \label{tab:race_breaches} 
\begin{tabular}{@{\extracolsep{5pt}}lc} 
\\[-1.8ex]\hline 
\hline \\[-1.8ex] 
 & \multicolumn{1}{c}{\textit{Dependent variable:}} \\ 
\cline{2-2} 
\\[-1.8ex] & Number of Breaches \\ 
\hline \\[-1.8ex] 
 Black & .04 \\ 
  & (.12) \\ 
  Hispanic/Latino & $-$.62$^{***}$ \\ 
  & (.10) \\ 
  Asian & $-$.30 \\ 
  & (.23) \\ 
  Native American & $-$.16 \\ 
  & (.36) \\ 
  Middle Eastern & $-$.46$^{**}$ \\ 
  & (.24) \\ 
  Mixed Race & $-$.67$^{**}$ \\ 
  & (.29) \\ 
  Other & $-$.20 \\ 
  & (.73) \\ 
  Constant & 3.12$^{***}$ \\ 
  & (.05) \\ 
 \hline \\[-1.8ex] 
Observations & 5,000 \\ 
R$^{2}$ & .01 \\ 
Adjusted R$^{2}$ & .01 \\ 
\hline 
\hline \\[-1.8ex] 
\textit{Note:}  & \multicolumn{1}{r}{$^{*}$p$<$0.1; $^{**}$p$<$0.05; $^{***}$p$<$0.01} \\ 
\end{tabular} 
\end{table} 

\clearpage

\begin{table}[!htbp] \centering 
  \caption{Number of Breaches by Sex} 
  \label{tab:sex_breaches} 
\begin{tabular}{@{\extracolsep{5pt}}lc} 
\\[-1.8ex]\hline 
\hline \\[-1.8ex] 
 & \multicolumn{1}{c}{\textit{Dependent variable:}} \\ 
\cline{2-2} 
\\[-1.8ex] & Number of Breaches \\ 
\hline \\[-1.8ex] 
 Male & $-$.35$^{***}$ \\ 
  & (.07) \\ 
  Constant & 3.17$^{***}$ \\ 
  & (.05) \\ 
 \hline \\[-1.8ex] 
Observations & 5,000 \\ 
R$^{2}$ & .004 \\ 
Adjusted R$^{2}$ & .004 \\ 
\hline 
\hline \\[-1.8ex] 
\textit{Note:}  & \multicolumn{1}{r}{$^{*}$p$<$0.1; $^{**}$p$<$0.05; $^{***}$p$<$0.01} \\ 
\end{tabular} 
\end{table} 

\clearpage

\begin{table}[!htbp] \centering 
  \caption{Number of Breaches by Education} 
  \label{tab:educ_breaches} 
\begin{tabular}{@{\extracolsep{5pt}}lc} 
\\[-1.8ex]\hline 
\hline \\[-1.8ex] 
 & \multicolumn{1}{c}{\textit{Dependent variable:}} \\ 
\cline{2-2} 
\\[-1.8ex] & Number of Breaches \\ 
\hline \\[-1.8ex] 
 HS Grad. & .54$^{***}$ \\ 
  & (.16) \\ 
  Some College & .69$^{***}$ \\ 
  & (.16) \\ 
  2-year College Degree & .72$^{***}$ \\ 
  & (.18) \\ 
  4-year College Degree & .87$^{***}$ \\ 
  & (.17) \\ 
  Postgrad Degree & .85$^{***}$ \\ 
  & (.18) \\ 
  Constant & 2.35$^{***}$ \\ 
  & (.14) \\ 
 \hline \\[-1.8ex] 
Observations & 5,000 \\ 
R$^{2}$ & .01 \\ 
Adjusted R$^{2}$ & .01 \\ 
\hline 
\hline \\[-1.8ex] 
\textit{Note:}  & \multicolumn{1}{r}{$^{*}$p$<$0.1; $^{**}$p$<$0.05; $^{***}$p$<$0.01} \\ 
\end{tabular} 
\end{table}

\clearpage

\begin{table}[!htbp] \centering 
  \caption{Number of Breaches by Age} 
  \label{tab:age_breaches} 
\begin{tabular}{@{\extracolsep{5pt}}lc} 
\\[-1.8ex]\hline 
\hline \\[-1.8ex] 
 & \multicolumn{1}{c}{\textit{Dependent variable:}} \\ 
\cline{2-2} 
\\[-1.8ex] & Number of Breaches \\ 
\hline \\[-1.8ex] 
 ns(age, 2)1 & 2.23$^{***}$ \\ 
  & (.20) \\ 
  ns(age, 2)2 & $-$.90$^{***}$ \\ 
  & (.21) \\ 
  Constant & 2.01$^{***}$ \\ 
  & (.09) \\ 
 \hline \\[-1.8ex] 
Observations & 5,000 \\ 
R$^{2}$ & .03 \\ 
Adjusted R$^{2}$ & .03 \\ 
\hline 
\hline \\[-1.8ex] 
\textit{Note:}  & \multicolumn{1}{r}{$^{*}$p$<$0.1; $^{**}$p$<$0.05; $^{***}$p$<$0.01} \\ 
\end{tabular} 
\end{table}

\clearpage
\subsection{Figures}
\label{fig:si_figs}

\begin{figure}[H]
  \centering
  \captionsetup{font={small,it}}
   \caption{Relationship Between Race and Number of Breaches
  \label{fig:race_breaches}}
    \includegraphics[scale=.75]{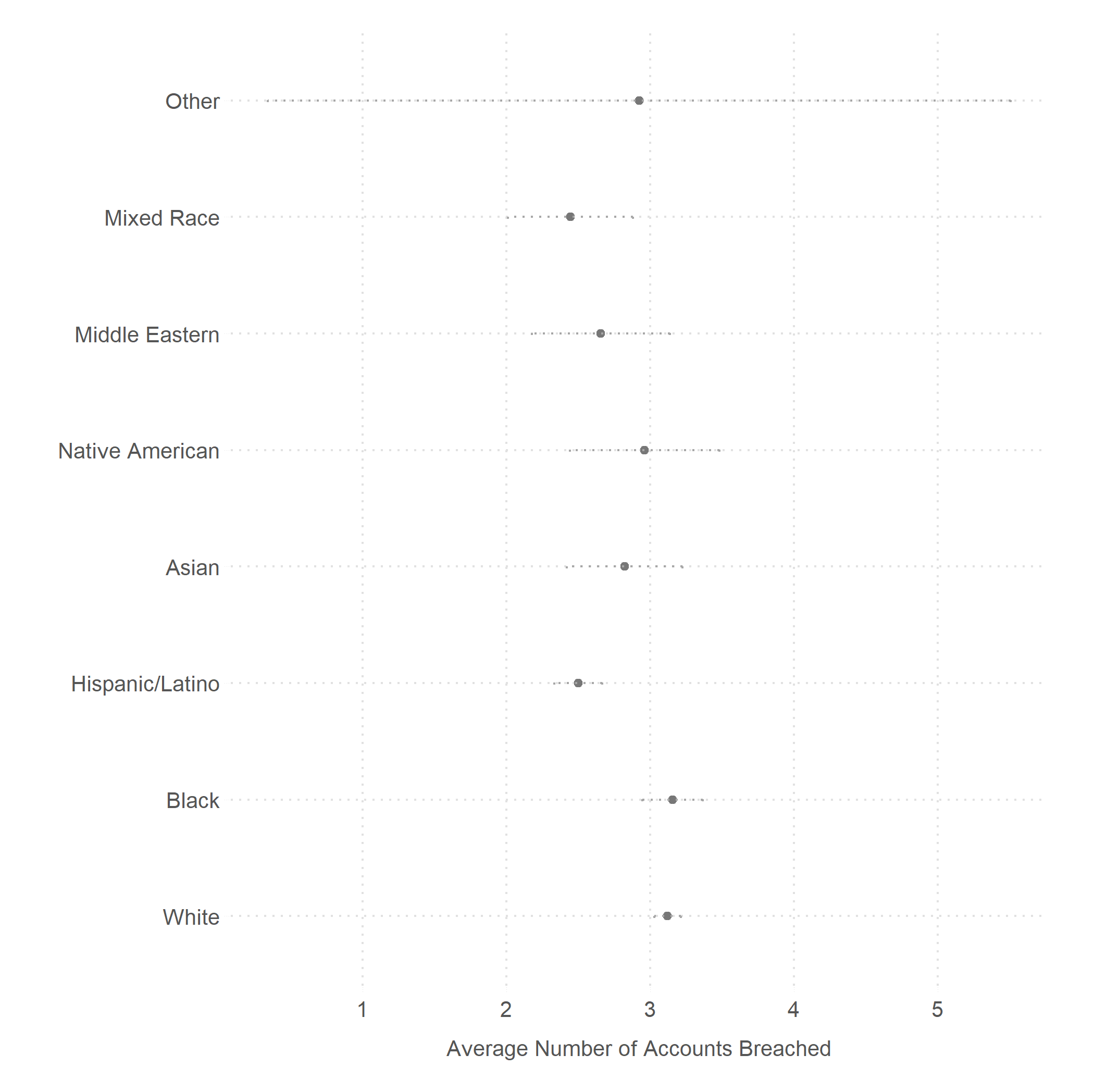}
\end{figure}
\clearpage

\begin{figure}[H]
  \centering
  \captionsetup{font={small,it}}
   \caption{Relationship Between Sex and Number of Breaches  
  \label{fig:sex_breaches}}
    \includegraphics[scale=.75]{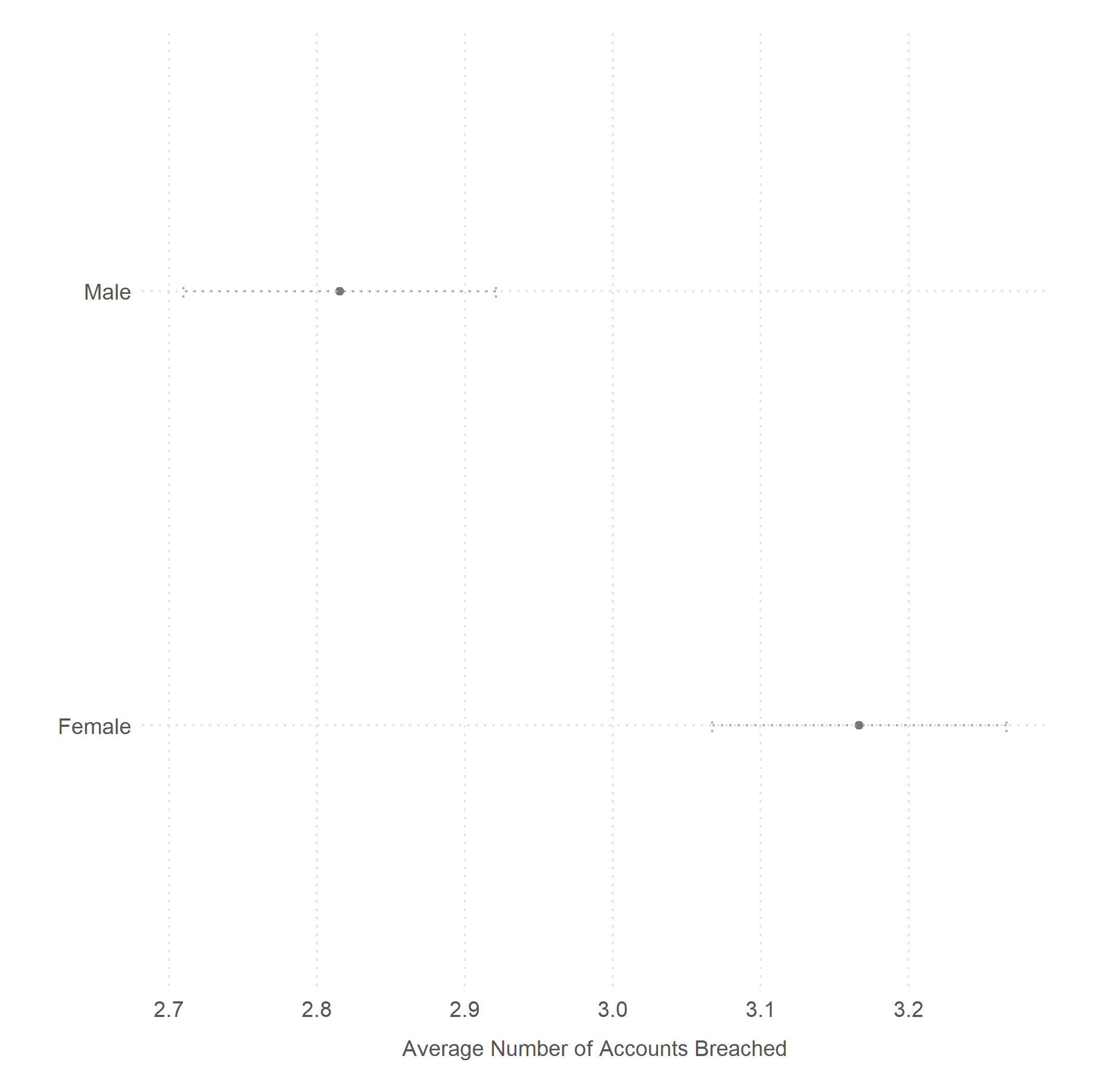}
\end{figure}
\clearpage

\begin{figure}[H]
  \centering
  \captionsetup{font={small,it}}
   \caption{Relationship Between Education and Number of Breaches 
  \label{fig:educ_breaches}}
    \includegraphics[scale=.75]{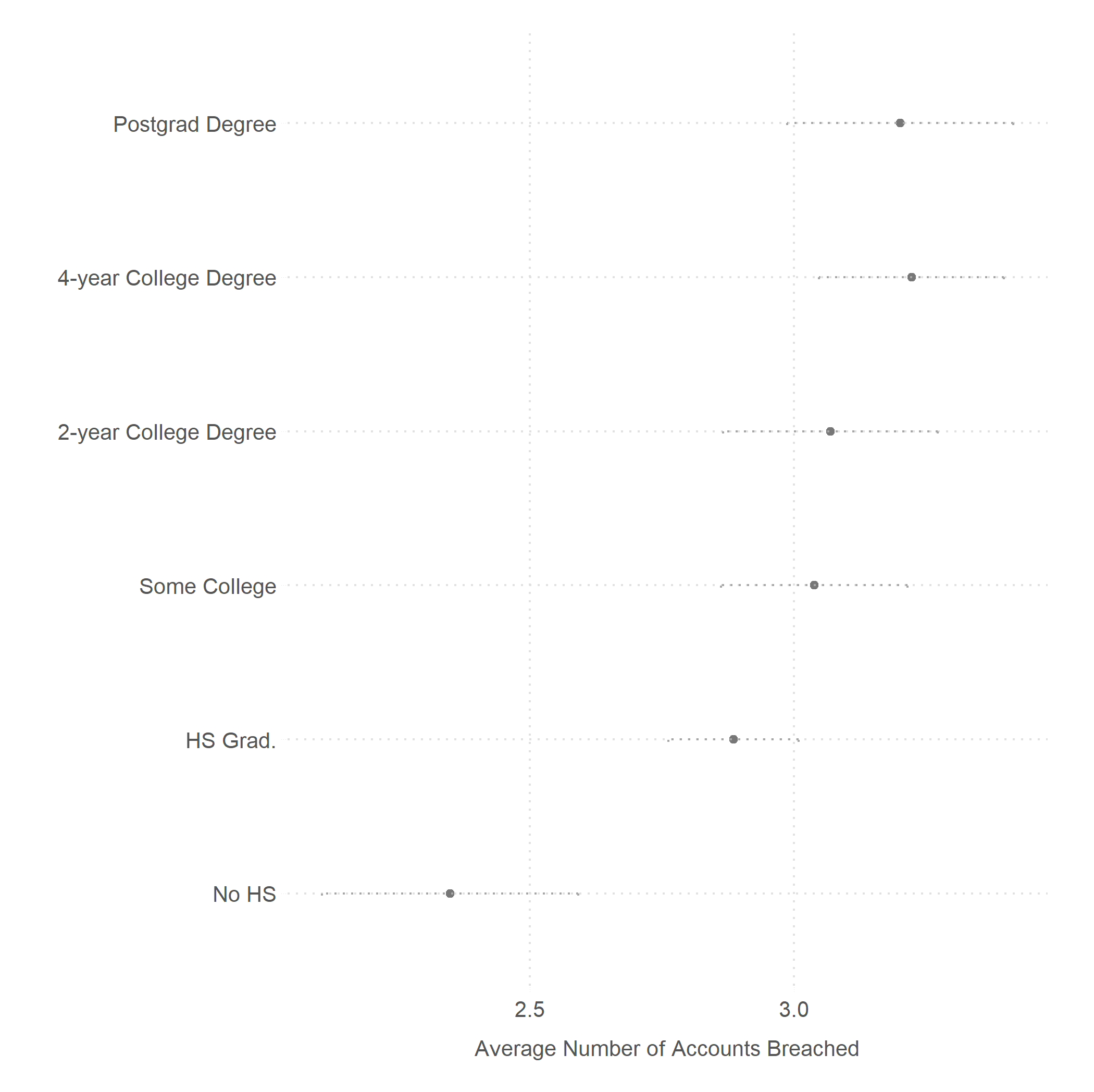}
\end{figure}
\clearpage

\begin{figure}[H]
  \centering
  \captionsetup{font={small,it}}
   \caption{Relationship Between Age and Number of Breaches  
  \label{fig:age_breaches}}
    \includegraphics[scale=.75]{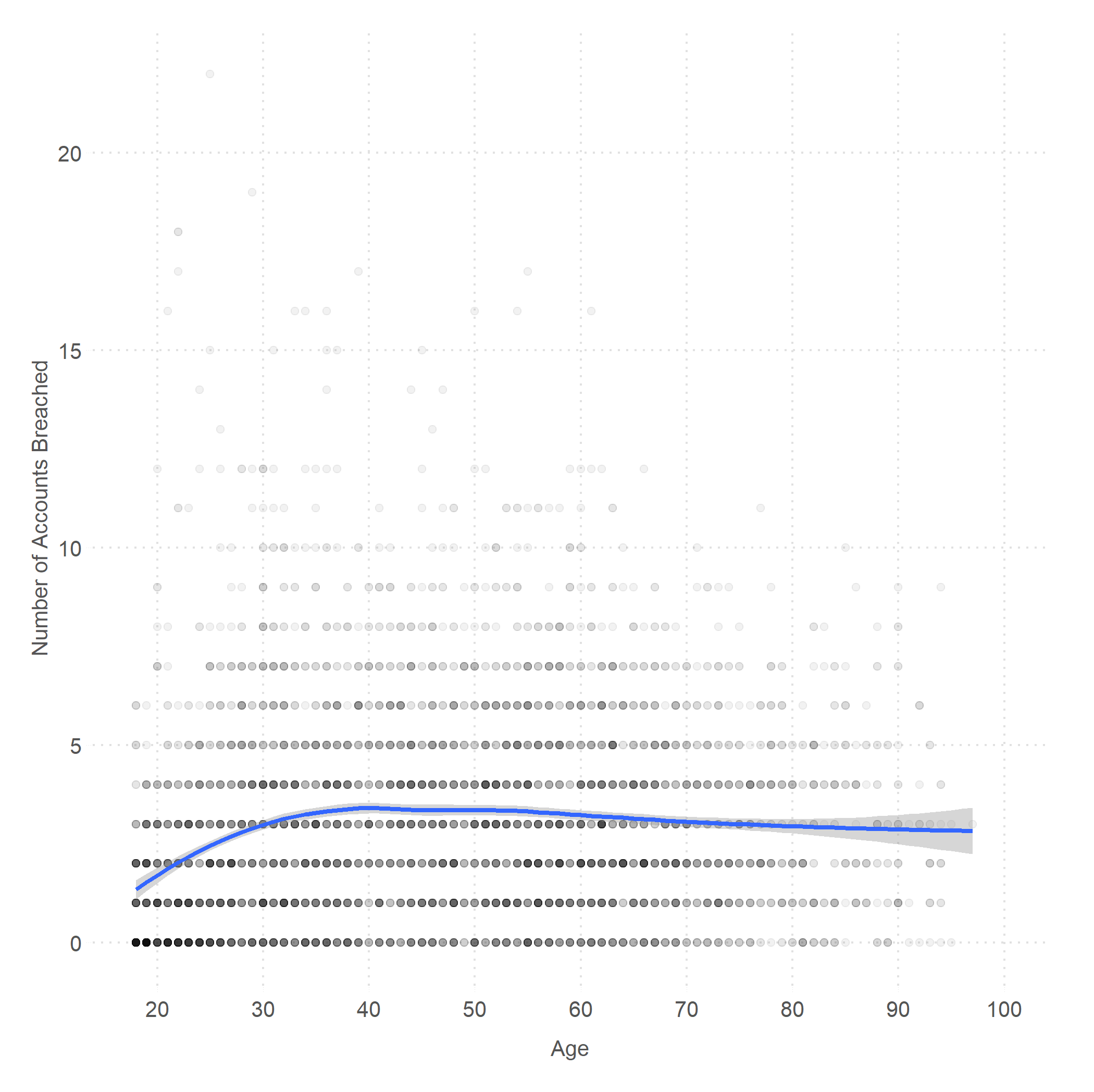}
\end{figure}

\end{document}